# Unusual topological polar texture in moiré ferroelectrics

Yuhao Li[1,2], Yuanhao Wei[3], Ruiping Guo[4,5], Yifei Wang[4], Hanhao Zhang[1], Takashi Taniguchi[6], Kenji Watanabe[7], Yan Shi[3], Yi Shi[1,*], Chong Wang[4,*], Zaiyao Fei[1,2,*]

1. National Laboratory of Solid-State Microstructures, School of Electronic Science and Engineering and Collaborative Innovation Center of Advanced Microstructures, Nanjing University, Nanjing 210093, Jiangsu, China
2. National Key Laboratory of Spintronics, Nanjing University, Suzhou 215163, Jiangsu, China
3. State Key Laboratory of Mechanics and Control for Aerospace Structures, Nanjing University of Aeronautics and Astronautics, Nanjing 210016, Jiangsu, China
4. State Key Laboratory of Low Dimensional Quantum Physics and Department of Physics, Tsinghua University, Beijing 100084, China
5. Institute for Advanced Study, Tsinghua University, Beijing 100084, China
6. Research Center for Materials Nanoarchitectonics, National Institute for Materials Science, 1-1 Namiki, Tsukuba 305-0044, Japan
7. Research Center for Electronic and Optical Materials, National Institute for Materials Science, 1-1 Namiki, Tsukuba 305-0044, Japan

*Correspondence to: zyfei@nju.edu.cn, chongwang@mail.tsinghua.edu.cn and yshi@nju.edu.cn

## Abstract

Topological polar textures in ferroelectrics have attracted significant interest for their potential for energy-efficient and high-density data storage and processing. Among these, polar merons and antimerons are predicted in strained and twisted bilayers of inversion symmetry broken systems. However, experimental observation of these polar textures within twisted two-dimensional van der Waals (2D vdW) materials remains challenging. Here, we utilize vector piezoresponse force microscopy (PFM) to reconstruct the polarization fields in R-type marginally twisted hexagonal boron nitride (hBN). We observe alternating out-of-plane (OOP) polarizations at domain regions and in-plane (IP) vortex-like polarization patterns along domain walls (DWs), indicative of a network of polar merons and antimerons. Notably, the OOP polarization exhibits three polarity reversals across a DW. Similar polar textures are identified in marginally twisted $MoSe_2$ and $WSe_2$ homobilayers. Our theoretical simulations attribute these unusual polarization reversals near the DWs to the competition between moiré ferroelectricity and piezoelectricity. These results provide experimental evidence of complex polar textures in moiré ferroelectrics, offering new insights into the electronic band topology in twisted transition metal dichalcogenides (TMDCs).

## Introduction

The intricate interplay of various degrees of freedom, such as lattice, charge, spin and orbit in condensed matter can give rise to diverse topological structures in real space[1-3]. For instance, the Dzyaloshinsky-Moriya spin-orbit interaction in magnetic thin films can foster topological structures like magnetic skyrmions[4,5] (winding number $N = \pm 1$) and merons[6] ($N = \pm 1/2$). A meron can be viewed as half a skyrmion with spins in the core area pointing OOP, gradually evolving into in-plane vortex-like pattern at the boundary. The electric counterpart, topological polar textures, have also been demonstrated but only in a few ferroelectric systems, such as oxide superlattices[2,7-11].

Recently, interfacial ferroelectricity has been discovered in noncentrosymmetric 2D vdW materials through specific stacking arrangements or interlayer twisting[12-17]. This new ferroelectric system provides unique opportunities for exploring and manipulating topological polar textures as well as exotic electronic states[18-26]. In particular, for R-type twisted hBN and TMDC homobilayers, alternating OOP ferroelectric polarization emerges due to inversion symmetry breaking, a phenomenon known as moiré ferroelectricity[27]. On the other hand, since the individual layer lacks inversion symmetry, the piezoelectric effect may also contribute to electric polarization in moiré ferroelectrics, especially near the DWs where strain is confined to. In experiment, OOP and IP polarizations have been observed in moiré ferroelectrics within the domains and near the DWs[28,29], respectively. However, the origin of the electric polarization and its evolution in these systems remain poorly understood.

In this study, we methodically investigate the OOP and IP electromechanical responses of marginally twisted double trilayer hBN, bilayer $MoSe_2$ and bilayer $WSe_2$ by vector PFM. Within the moiré domains, only OOP polarizations are detected, exhibiting opposite polarities in adjacent domains, consistent with moiré ferroelectricity. Near the DWs, both OOP and IP polarizations are observed. The IP components are found to align along the DWs through angular-dependent lateral PFM (LPFM) measurement, forming vortex-like structures. Of particular interest is the distribution of OOP components near the DWs. By suppressing the cantilever buckling effect induced by IP polarization during vertical PFM (VPFM) measurement, we obtain the intrinsic OOP polarization field near the DWs. The OOP polarization in twisted hBN reverses three times across a DW from AB to BA stackings, instead of a monotonical transition. The unusual polar textures are also observed in twisted $MoSe_2$ and $WSe_2$ homobilayers, revealing a competitive interplay between ferroelectricity and piezoelectricity in moiré ferroelectrics.

## Results

We first look at twisted hBN samples with a tiny twist angle of below 0.05°, which are comprised of exposed twisted hBN sitting atop thick h-BN substrates (see Methods and Fig. S1 for fabrication details). Figure 1a shows the atomic arrangements for AB, BA and saddle-point (SP) stackings of a R-type twisted bilayer hBN moiré superlattices (Fig. 1b). The yellow and purple arrows indicate the orientations of electric dipoles at the interface of AB and BA stackings, respectively. Due to an in-plane two-fold rotational symmetry, the SP stacking has vanishing OOP electric dipoles.

Figure 1c shows the Kelvin probe force microscopy (KPFM) image of a twisted double trilayer hBN sample, which gives negative and positive surface potential for the commensurate AB and BA stacking domains, respectively. While KPFM effectively images surface potential, it primarily reflects the OOP polarization and cannot measure the IP polarization near the DWs. Furthermore, the DWs separating the AB and BA stacking domains are on the order of 10 nm wide, resulting from atomic reconstruction, which exceeds the resolution limit of KPFM. To reconstruct the full polarization field with higher resolution, we switch to vector PFM for the rest of this study. During the measurements, the cantilevers are always aligned along the horizonal axis, unless otherwise specified. Typically, the VPFM signal (vertical deformation) is associated with the OOP component of the electric polarization, whereas of the LPFM signal (lateral deformation) is associated with the IP component.

Figure 1d&e show the decoupled VPFM phase and amplitude images of the same sample (see Figs. S2a&b and Supplementary Note 1 for the raw images and discussions of the decoupling processes). A 180° phase contrast is observed between the AB and BA stacking domains, while their amplitudes are comparable. LPFM images of the same area show vanishing IP signals within the AB and BA domains (Fig. S3). These suggest that the adjacent stacking domains are oppositely polarized OOP, consistent with the KPFM image in Fig. 1c. Surprisingly, for different moiré DWs, the VPFM signals seem to depend on the relative orientation between the cantilever and the DW, which is also illustrated in Fig. 2d and Supplementary Note 2. This contradicts the expected three-fold rotational symmetry for the OOP polarization. Moreover, across DWs that are nearly perpendicular to the cantilever, the in-phase VPFM signals, $-A \cdot \cos(\theta)$ (the minus sign aligns the signal with the polarization direction), exhibit extra polarity reversals, as exampled in the line cut

of DW1 (Fig. 1f).

To understand the DW-specific VPFM signals, we perform a series of vector PFM measurements while rotating the sample. The selected measurement schemes (top insets) and decoupled LPFM images (middle and bottom insets) are shown in Fig. 2a-c, where ψ measures the rotation of the sample. In the measurement schemes, we intentionally rotate the cantilever to improve visualization. The normalized in-phase LPFM signal, $-A_n \cdot \cos(\theta)$, as a function of the cantilever-DW angle (φ) is plotted in Fig. 2d (see Fig. S4 for the normalization process), which can be fitted to a sinusoidal function (purple line). Since the LPFM detects the component of the IP polarization perpendicular to the cantilever[30], the sinusoidal relation implies that the IP polarization near the DWs are along the DWs, thus forming clockwise and anti-clockwise vortex-like patterns, similar to the results obtained in graphene moiré patterns[31-33]. On the other hand, the normalized in-phase VPFM signal after averaging on each DW can be fitted into a cosine curve (orange line, the angle-dependent VPFM images can be found in Fig. S5). The observed 90° shift suggests that the averaged VPFM signals can be associate with a component of the IP polarization that lies along the cantilever direction, which is known as IP polarization induced cantilever buckling effect[34,35]. Details of the buckling effect in our measurements can be found in Fig. S6 and Supplementary Note 2. In Fig. S7&S8, we provide angle-dependent vector PFM measurements on another twisted hBN sample, which reveals similar results.

To suppress the buckling effect, we can align the DWs perpendicular to the cantilever, as for the highlighted DW (DW1) in Fig. 1d&e. VPFM images of the same sample while DW2&3 are aligned perpendicular to the cantilever are provided in Fig. S5. In this case, the in-phase VPFM signal in Fig. 1g(i) is assigned to the OOP polarization. Three polarity reversals are observed across the DW for the OOP polarization, marking the key finding of this work. At the center of the DW (SP stacking), the OOP polarization vanishes, consistent with the symmetry analysis above. We also plot the moiré ferroelectric contribution to the VPFM signal in Fig. 1g(ii)[36]. The remaining VPFM signal (Fig. 1g(iii)) is attributed to piezoelectric effect[37], as corroborated by the DFT calculations presented later. We note the piezoelectric contribution to the polarization has the opposite sign to that of the moiré ferroelectricity, indicating a competitive interplay between the two within the DWs. We also realize the relative amplitude of the buckling effect varies with several factors, such as the position of the laser spot (see Supplementary Note 2 and Fig. S9 for more information).

Alternatively, the intrinsic OOP polarization can be extrapolated from symmetry analysis. As AB and BA stackings are related by an OOP mirror operation, the OOP polarization is antisymmetric with respect to the SP stacking, while the IP polarization is symmetric. Figure 2e&f show the antisymmetrized VPFM images, where three polarity reversals are observed across nearly all DWs, restoring the three-fold rotational symmetry.

We also perform vector PFM measurements on R-type marginally twisted $MoSe_2$ and $WSe_2$ homobilayers. Similar to twisted hBN, the commensurate AB and BA stacking domains are also polarized in the OOP directions, whereas the SP stacking domains are nonpolar. Figure 3a&b show VPFM images of a marginally twisted bilayer $MoSe_2$ sample. Clearly, the AB and BA stacking domains also exhibit the expected 180º phase contrast, with VPFM amplitudes of ~10 pm, nearly half of that observed in twisted hBN. Near the DWs, the features are less pronounced but still distinguishable. Three polarity reversals are observed in the VPFM signal (Fig. 3c), qualitatively consistent with that in twisted hBN. The VPFM images of a twisted $WSe_2$ sample and LPFM images of twisted $MoSe_2$ are provided in Fig. S10&S11.

With collected information of both IP and OOP polarizations, we sketch the polar texture of twisted hBN or TMDC homobilayer in Fig. 3d. The yellow and purple shaded triangles denote the commensurate AB and BA stacking domains, where the polarization is purely OOP. In the DW regions, IP polarization emerges, as indicated by the red arrows. Across the DW (denoted by the black arrowed line), the OOP polarization first reverses, reaching a peak value before decreasing in amplitude and vanishing at the DW center. The OOP polarization profile on the other side of the DW is an antisymmetric copy. The polar texture across the DW can also be visualized on a ferroelectric Bloch sphere, as shown in Fig. 3f. For AB stacking, the polarization vector points to the south pole. Across the DW, it rotates to equator first and tilts toward the north pole during OOP polarization reversal. It then returns to the equator at the DW center, where the OOP polarization vanishes.

To illustrate the competition between moiré ferroelectricity and piezoelectricity in twisted hBN or $MoSe_2$, we conducted relaxation simulations by elastic-theory-based continuum model for twisted bilayer $MoSe_2$ at a twist angle of 0.2° (See Methods for details). The relaxed structure, IP and OOP polarization maps are shown in Fig. 4. After relaxation, large domains with uniform AB and BA stackings emerge, as schematically depicted in Fig. 4a. Moiré ferroelectricity generates both

IP polarization along the DWs and OOP polarization within the AB and BA domains (Fig. 4b&c). Within the DW regions, piezoelectricity induces significant OOP polarization (Fig. 4d), opposite in sign to the ferroelectric polarization. Consequently, ferroelectric OOP polarization dominates in the domains, while in the DWs, the stronger piezoelectric OOP polarization reverses the sign of the polarization across the DWs (Fig. 4e&f). We also calculate the winding numbers of the unit polarization field for the extended BA and AB domains (including half of the domain walls), which are

$$N = \frac{1}{4\pi} \int \hat{P} \cdot (\partial_x \hat{P} \times \partial_y \hat{P}) ds = \pm 1/2$$

This suggests they are polar merons and antimerons, as schematically shown in Fig. 3e.

**Discussion**

The unusual polar textures we observed in marginally twisted hBN and TMDC moiré superlattices are reminiscent of the polarization charge distributions in 1.25° twisted bilayer $MoTe_2$ or $WSe_2$ moiré superlattice[37]. Limited by the spatial resolution of PFM (sub 10 nm)[38], transitions of moiré potential maxima at larger twist angles are unfortunately beyond the reach of this technique. Future studies employing more advanced imaging techniques, such as high-resolution transmission electron microscopy or scanning tunneling microscopy could provide a more comprehensive understanding of the correlation between real-space polar textures and the band topology in these systems.

Overall, our study presents the first experimental evidence of complex topological polar textures in marginally twisted hBN and TMDC homobilayers, highlighting the unique interplay between moiré ferroelectricity and piezoelectricity. The ability to modulate the polar texture through stacking configurations[39-41], twist angle, or interlayer sliding offers exciting possibilities for engineering novel topological and electronic states. This opens pathways for integrating moiré ferroelectrics into nanoscale devices, including topological memory elements and reconfigurable logic gates.

**Methods**

**Sample fabrication.** All devices were fabricated using a modified 'tear-and-stack' method, similar to Ref.[42]. Typical optical images of twisted hBN and TMDC are presented in Fig. S1.

**Atomic force microscopy measurements.** Three AFM modes were adopted in this work, i.e., VPFM, LPFM, and KPFM, all performed on an Oxford Instruments Asylum Research MFP-3D Origin AFM. ASYELEC-01-R2 probes coated with 5-nm Ti and 20-nm Ir were used in all modes with a spring constant of 2.8 N/m. Typical free resonance frequency is ~75 kHz. VPFM and LPFM contact resonance frequency are ~300 kHz and ~780 kHz, respectively. In KPFM measurement, the dual pass amplitude modulation is employed to capture the topography and surface potential difference between tip and sample, respectively. To minimize the potential shift, a zero-order flattening process is applied to the raw surface potential data.

**Data processing.** The in-phase LPFM and VPFM signals at DWs are collected and Gaussian fittings are performed on them. The mean values and standard deviations of the fittings for LPFM correspond to the projected IP deformations and corresponding error bars, and the VPFM results are the buckling effect of IP deformations. By averaging the VPFM signal for each DW, the contribution of $P_z$ is excluded, leaving only the contribution of IP polarization induced buckling effect.

**Theoretical modelling of local polarization.** To simulate the structural relaxation and piezoelectric polarization shown in Fig. 4, we employ an elastic-theory-based method described in Ref.[43]. This approach decomposes the total energy into elastic and stacking energies, optimizing the continuum atomic displacement field to minimize the total energy. For the elastic energy calculations, we use a bulk modulus $K = 40521 \text{ meV/u.c.}$ and a shear modulus $G = 26464 \text{ meV/u.c.}$ for monolayer MoSe$_2$, as reported in Ref.[44]. The stacking energy is characterized using the generalized stacking fault energy (GSFE) for parallel bilayer MoSe$_2$ from Ref.[45]. From the relaxation results, we separately compute the piezoelectric polarization and the ferroelectric polarization. The piezoelectric charge density is computed with the piezoelectric coefficient $e_{11} \approx 435 \ pC/m$ from Ref.[46]. The piezoelectric charge density is assumed to be completely accounted for by the piezoelectric OOP polarization. The ferroelectric polarization pattern in the real space is obtained by mapping the real space to different stacking configurations. By symmetry analysis, up to the first harmonics, the IP and OOP ferroelectric polarization for different stacking configurations have the

form

$$P_{xy} = c_{xy} \cdot \begin{pmatrix} 2\sqrt{3} \sin\left(\frac{\sqrt{3}x}{2}\right) \sin\left(\frac{y}{2}\right) \\ 2\cos(y) - 2\cos\left(\frac{\sqrt{3}x}{2}\right) \cos\left(\frac{y}{2}\right) \end{pmatrix},$$

$$P_z = c_z \cdot \left(2\sin(y) - 4\cos\left(\frac{\sqrt{3}x}{2}\right) \sin\left(\frac{y}{2}\right)\right),$$

where the lattice vectors are scaled to be $2\pi\,(\frac{1}{2}, \frac{\sqrt{3}}{2})$ and $2\pi\,(-\frac{1}{2}, \frac{\sqrt{3}}{2})$. Based on the modern theory of polarization[47], we compute the polarization by density functional theory (DFT) and the coefficients are fitted to be $c_{xy} = 0.324\,\text{pC/m}$, $c_z = 0.0717\,\text{pC/m}$.

**Density functional theory calculations.** DFT calculations were performed using the VASP package[48], employing the projector augmented wave pseudopotential[49,50] and the Perdew-Burke-Ernzerhof (PBE) exchange-correlation functional[51]. The relaxation calculations were performed using a $19 \times 19 \times 1$ Gamma-centered k-point grid, an energy cutoff of 800 eV, the DFT-ulg method[52] for van der Waals correction. The convergence criteria for atomic forces in structural relaxation is $10^{-2}$ eV/Å.

**Acknowledgements**

This work is supported by the National Key Research and Development Program of China (2021YFA0715600), National Natural Science Foundation of China (12274222 and 12404217), the National Science Foundation of Jiangsu Province (BK20220756) and the Fundamental Research Funds for the Central Universities (2024300420). Yan Shi is supported by National Natural Science Foundation of China (12072150) and National Natural Science Foundation of China for Creative Research Groups (51921003). C.W., R.G. and Y.W. are supported by the startup grant from Tsinghua University. C.W., R.G. and Y.W. acknowledge the discussion with Qiyun Xu and Bozhong Zhang. K.W. and T.T. acknowledge support from the JSPS KAKENHI (21H05233 and 23H02052), the CREST (JPMJCR24A5), JST and World Premier International Research Center Initiative (WPI), MEXT, Japan.

**Author contributions**

Z.F. and Yi Shi supervised the project. H.Z., Y.W. and Y.L. fabricated the devices. Y.L. and Y.W. performed the AFM measurements, R.G., Y.W. and C.W. performed the theoretical modelling. T.T.

and K.W. provided the bulk BN crystals. Z.F., Y.L., C.W. and Y.W. analyzed the data. Z.F., Y.L., and C.W. wrote the paper with inputs from all authors.

**Competing interests**

The authors declare no competing interests.

**Additional information**

Supplementary information is available for this paper on the website.

Correspondence and requests for materials should be addressed to Z.F.

**Data availability**

The data that support the findings of this study are available from the corresponding author upon reasonable request.

**References:**

1   Hwang, H. Y. *et al.* Emergent phenomena at oxide interfaces. Nature Materials 11, 103-113 (2012).
2   Wang, Y. J. *et al.* Polar meron lattice in strained oxide ferroelectrics. Nature Materials 19, 881-886 (2020).
3   Junquera, J. *et al.* Topological phases in polar oxide nanostructures. Reviews of Modern Physics 95, 025001 (2023).
4   Mühlbauer, S. *et al.* Skyrmion Lattice in a Chiral Magnet. Science 323, 915-919 (2009).
5   Nagaosa, N. & Tokura, Y. Topological properties and dynamics of magnetic skyrmions. Nature Nanotechnology 8, 899-911 (2013).
6   Lin, S.-Z., Saxena, A. & Batista, C. D. Skyrmion fractionalization and merons in chiral magnets with easy-plane anisotropy. Physical Review B 91, 224407 (2015).
7   Yadav, A. K. *et al.* Observation of polar vortices in oxide superlattices. Nature 530, 198-201 (2016).
8   Das, S. *et al.* Observation of room-temperature polar skyrmions. Nature 568, 368-372 (2019).
9   Han, L. *et al.* High-density switchable skyrmion-like polar nanodomains integrated on silicon. Nature 603, 63-67 (2022).
10  Shao, Y.-T. *et al.* Emergent chirality in a polar meron to skyrmion phase transition. Nature Communications 14, 1355 (2023).
11  Chen, H. *et al.* Highly tunable skyrmion-like polar nanodomains for high-density ferroelectric hard disks. Applied Physics Reviews 11 (2024).
12  Li, L. & Wu, M. Binary Compound Bilayer and Multilayer with Vertical Polarizations: Two-Dimensional Ferroelectrics, Multiferroics, and Nanogenerators. ACS Nano 11, 6382-6388 (2017).
13  Vizner Stern, M. *et al.* Interfacial ferroelectricity by van der Waals sliding. Science 372, 1462-1466 (2021).

14  Yasuda, K., Wang, X., Watanabe, K., Taniguchi, T. & Jarillo-Herrero, P. Stacking-engineered ferroelectricity in bilayer boron nitride. Science 372, 1458-1462 (2021).

15  Rogée, L. *et al.* Ferroelectricity in untwisted heterobilayers of transition metal dichalcogenides. Science 376, 973-978 (2022).

16  Weston, A. *et al.* Interfacial ferroelectricity in marginally twisted 2D semiconductors. Nat. Nanotechnol. 17, 390-395 (2022).

17  Ko, K. *et al.* Operando electron microscopy investigation of polar domain dynamics in twisted van der Waals homobilayers. Nat. Mater. (2023).

18  Tsang, C. S. *et al.* Polar and quasicrystal vortex observed in twisted-bilayer molybdenum disulfide. Science 386, 198-205 (2024).

19  Seyler, K. L. *et al.* Signatures of moire-trapped valley excitons in MoSe2/WSe2 heterobilayers. Nature (2019).

20  Tran, K. *et al.* Evidence for moiré excitons in van der Waals heterostructures. Nature 567, 71-75 (2019).

21  Liu, E. *et al.* Signatures of moiré trions in WSe2/MoSe2 heterobilayers. Nature 594, 46-50 (2021).

22  Wang, X. *et al.* Moiré trions in MoSe2/WSe2 heterobilayers. Nature Nanotechnology 16, 1208-1213 (2021).

23  Li, T. *et al.* Quantum anomalous Hall effect from intertwined moiré bands. Nature 600, 641-646 (2021).

24  Kang, K. *et al.* Evidence of the fractional quantum spin Hall effect in moiré MoTe2. Nature 628, 522-526 (2024).

25  Park, H. *et al.* Observation of fractionally quantized anomalous Hall effect. Nature 622, 74-79 (2023).

26  Xia, Y. *et al.* Superconductivity in twisted bilayer WSe2. Nature (2024).

27  Wu, M. & Li, J. Sliding ferroelectricity in 2D van der Waals materials: Related physics and future opportunities. Proceedings of the National Academy of Sciences 118, e2115703118 (2021).

28  Wan, S. *et al.* Intertwined Flexoelectricity and Stacking Ferroelectricity in Marginally Twisted hBN Moiré Superlattice. Advanced Materials n/a, 2410563 (2024).

29  Wang, X. *et al.* Interfacial ferroelectricity in rhombohedral-stacked bilayer transition metal dichalcogenides. Nature Nanotechnology 17, 367–371 (2022).

30  Li, J., Li, J.-F., Yu, Q., Chen, Q. N. & Xie, S. Strain-based scanning probe microscopies for functional materials, biological structures, and electrochemical systems. J. Materiomics 1, 3-21 (2015).

31  McGilly, L. J. *et al.* Visualization of moire superlattices. Nat. Nanotechnol. 15, 580–584 (2020).

32  Li, Y. *et al.* Unraveling Strain Gradient Induced Electromechanical Coupling in Twisted Double Bilayer Graphene Moiré Superlattices. Adv. Mater. 33, 2105879 (2021).

33  Zhang, H. *et al.* Layer-Dependent Electromechanical Response in Twisted Graphene Moiré Superlattices. ACS Nano 18, 17570-17577 (2024).

34  Nath, R. *et al.* Effects of cantilever buckling on vector piezoresponse force microscopy imaging of ferroelectric domains in BiFeO3 nanostructures. Applied Physics Letters 96 (2010).

35   Alikin, D. O. *et al.* In-plane polarization contribution to the vertical piezoresponse force microscopy signal mediated by the cantilever "buckling". Applied Surface Science 543 (2021).

36   Bennett, D., Chaudhary, G., Slager, R.-J., Bousquet, E. & Ghosez, P. Polar meron-antimeron networks in strained and twisted bilayers. Nature Communications 14, 1629 (2023).

37   Zhang, X.-W. *et al.* Polarization-driven band topology evolution in twisted MoTe2 and WSe2. Nat. Commun. 15, 4223 (2024).

38   Kalinin, S. V. *et al.* Spatial resolution, information limit, and contrast transfer in piezoresponse force microscopy. Nanotechnology 17, 3400 (2006).

39   Deb, S. *et al.* Cumulative polarization in conductive interfacial ferroelectrics. Nature 612, 465–469 (2022).

40   Meng, P. *et al.* Sliding induced multiple polarization states in two-dimensional ferroelectrics. Nature Communications 13, 7696 (2022).

41   Kim, D. S. *et al.* Electrostatic moiré potential from twisted hexagonal boron nitride layers. Nature Materials 23, 65-70 (2024).

42   Wong, D. *et al.* Cascade of electronic transitions in magic-angle twisted bilayer graphene. Nature 582, 198-202 (2020).

43   Wang, C., Gao, Y., Lv, H., Xu, X. & Xiao, D. Stacking Domain Wall Magnons in Twisted van der Waals Magnets. Physical Review Letters 125, 247201 (2020).

44   Shabani, S. *et al.* Deep moiré potentials in twisted transition metal dichalcogenide bilayers. Nature Physics 17, 720-725 (2021).

45   Enaldiev, V. V., Zólyomi, V., Yelgel, C., Magorrian, S. J. & Fal'ko, V. I. Stacking Domains and Dislocation Networks in Marginally Twisted Bilayers of Transition Metal Dichalcogenides. Physical Review Letters 124, 206101 (2020).

46   Ma, W. *et al.* Synergetic contribution of enriched selenium vacancies and out-of-plane ferroelectric polarization in AB-stacked MoSe2 nanosheets as efficient piezocatalysts for TC degradation. New Journal of Chemistry 46, 4666-4676 (2022).

47   King-Smith, R. D. & Vanderbilt, D. Theory of polarization of crystalline solids. Physical Review B 47, 1651-1654 (1993).

48   G. Kresse & Furthmiiller, J. Efficiency of ab-initio total energy calculations for metals and semiconductors using a plane-wave basis set. Comput. Mater. Sci. 6, 15-50 (1996).

49   Blöchl, P. E. Projector augmented-wave method. Phys. Rev. B 50, 17953-17979 (1994).

50   Kresse, G. & Joubert, D. From ultrasoft pseudopotentials to the projector augmented-wave method. Phys. Rev. B 59, 1758-1775 (1999).

51   Perdew, J. P., Burke, K. & Ernzerhof, M. Generalized Gradient Approximation Made Simple. Phys. Rev. Lett. 77, 3865-3868 (1996).

52   Kim, H., Choi, J.-M. & Goddard, W. A., III. Universal Correction of Density Functional Theory to Include London Dispersion (up to Lr, Element 103). The Journal of Physical Chemistry Letters 3, 360-363 (2012).

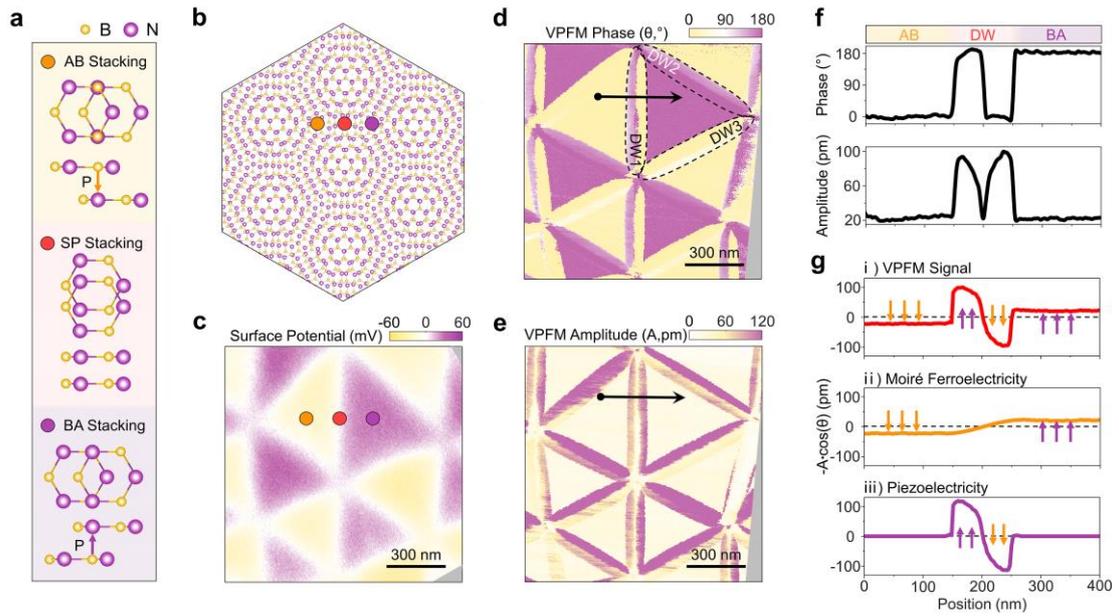

**Figure 1. KPFM and VPFM measurements on a marginally twisted hBN sample. a,** Schematics of AB, BA and SP stacking domains in a twisted hBN moiré superlattices. **b,** Schematic of a twisted hBN moiré superlattices. **c-e,** Surface potential, VPFM phase (θ) and VPFM amplitude (A) images of a twisted double trilayer hBN. **f,** Line cut of VPFM phase and amplitude along the black arrowed line in **d&e**. **g,** Line cuts of (i) measured in-phase VPFM signal (-A·cos(θ)), (ii) in-phase VPFM signal due to moiré ferroelectricity and (iii) in-phase VPFM signal due to piezoelectricity along the black arrowed lines in **d&e**.

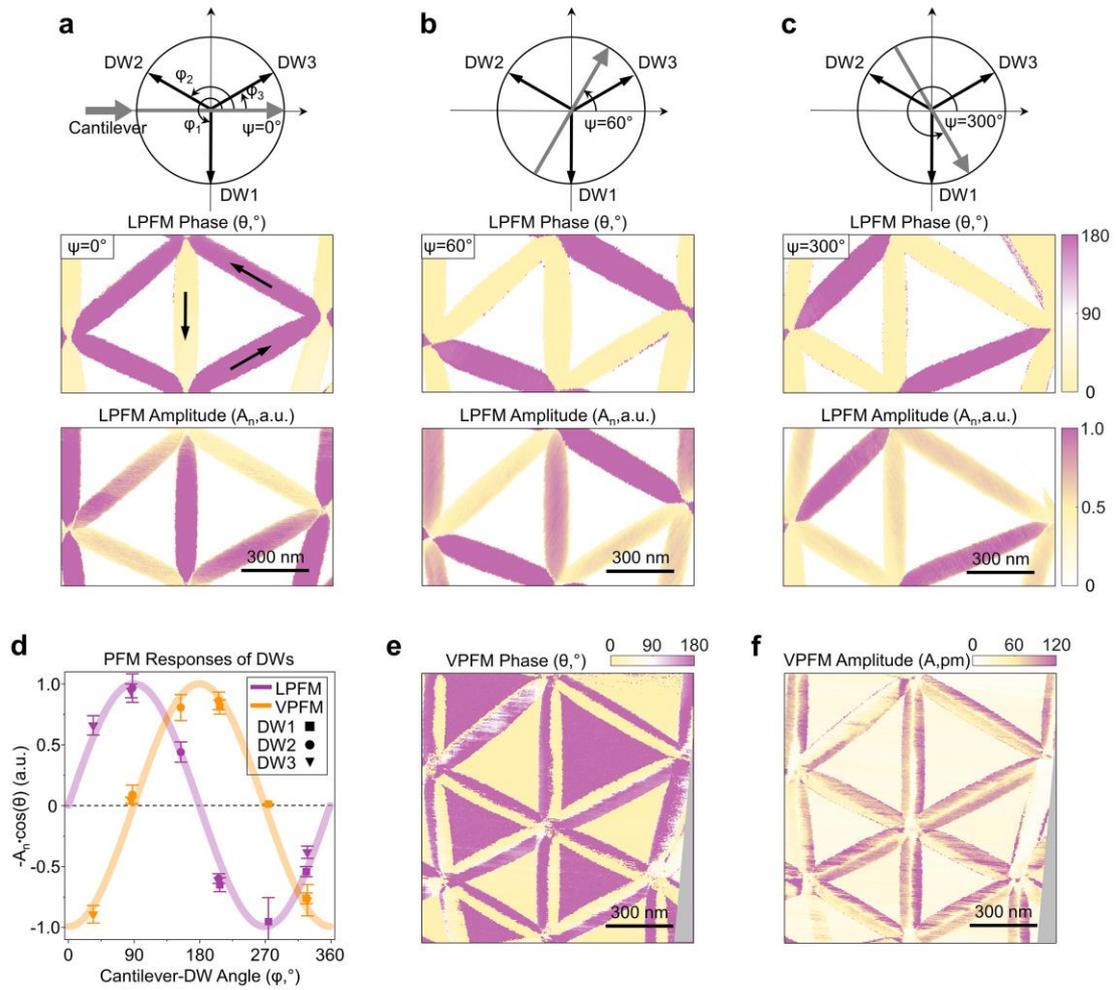

**Figure 2. Angle-dependent LPFM measurements and buckling effect-free VPFM images. a-c**, Selected measurement schemes (top insets) and corresponding LPFM phase and normalized LPFM amplitude images (middle and bottom insets). The cantilever-sample angle (ψ) are 0°, 60° and 300°, respectively. **d**, The normalized in-phase LPFM and VPFM signal ($-A_n \cdot \cos(\theta)$) as a function of the cantilever-DW angle (φ). The VPFM signals are shifted by 90° with respect to those of LPFM for all DWs. **e,f,** Buckling effect-free VPFM phase and amplitude images of Fig. 1d&e.

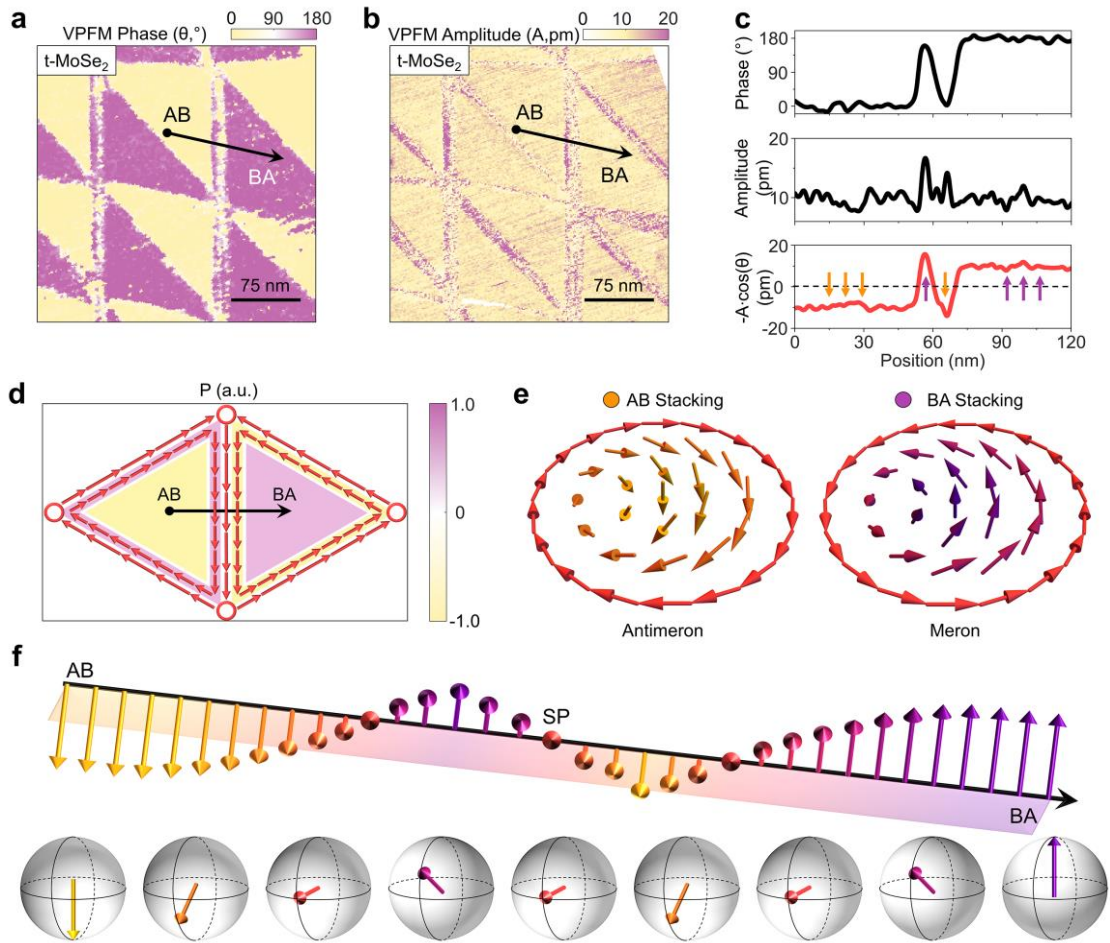

**Figure 3. VPFM measurement of a twisted bilayer MoSe$_2$ sample and evolution of the electric polarization. a,b,** VPFM phase (**a**) and amplitude (**b**) images of a twisted bilayer MoSe$_2$ sample. **c,** VPFM signals along the black arrowed lines in **a&b. d,** Schematic of the polar texture of the twisted hBN in Fig. 1&2. **e,** Polar meron and antimeron structures. **f,** Evolution of the electric polarization along the black arrowed line in **d** and the relative orientations on a ferroelectric Bloch sphere.

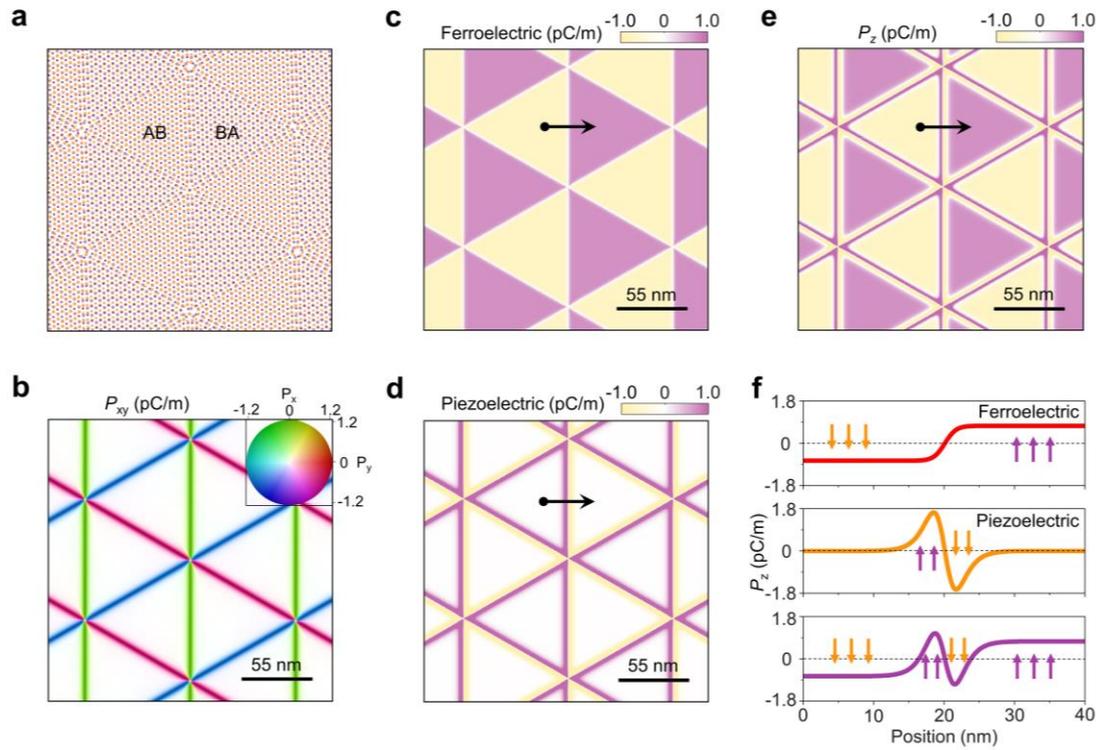

**Figure 4. Competition between moiré ferroelectricity and piezoelectricity in a twisted bilayer MoSe$_2$ moiré superlattice. a,** Atomic structure of a relaxed twisted bilayer MoSe$_2$ moiré superlattice. **b,** Calculated IP polarization map of twisted bilayer MoSe$_2$ moiré superlattices. **c-e,** Calculated OOP ferroelectric polarization, OOP piezoelectric polarization and total OOP polarization maps of twisted bilayer MoSe$_2$ moiré superlattices. **f,** Line cuts of the OOP polarizations in **c-e**.